# Calibration Method of the Monocular Omnidirectional Stereo Camera

**Ryota Kawamata** [1)]   **Keiichi Betsui** [1)]   **Kazuyoshi Yamazaki** [1)]   **Rei Sakakibara** [1)]   **Takeshi Shimano** [1)]

1) Center for Technology Innovation – Instrumentation, Hitachi, Ltd.,
1-280, Higashi-koigakubo, Kokubunji-shi, Tokyo, 185-8601, Japan (E-mail: ryota.kawamata.yv@hitachi.com)

**ABSTRACT**: Compact and low-cost devices are needed for autonomous driving to image and measure distances to objects 360-degree around. We have been developing an omnidirectional stereo camera exploiting two hyperbolic mirrors and a single set of a lens and sensor, which makes this camera compact and cost efficient. We establish a new calibration method for this camera considering higher-order radial distortion, detailed tangential distortion, an image sensor tilt, and a lens-mirror offset. Our method reduces the calibration error by 6.0 and 4.3 times for the upper- and lower-view images, respectively. The random error of the distance measurement is 4.9% and the systematic error is 5.7% up to objects 14 meters apart, which is improved almost nine times compared to the conventional method. The remaining distance errors is due to a degraded optical resolution of the prototype, which we plan to make further improvements as future work.

**KEY WORDS**: electronics and control, semiconductor camera, image processing / Stereo camera [E1]

## 1. Introduction

Autonomous cars need to perceive their surroundings, such as approaching cars and pedestrians. Thus, devices are desired to detect and measure distances to objects 360-degree around. Requirements of the devices are a large field of view (FoV), being able to detect objects, enough distance accuracy, and affordable cost.

Some studies and companies try to satisfy those requirements using a combination of LiDARs (Light Detection and Ranging) and cameras [e.g., 1, 2]. In this system, the LiDAR offers a large FoV and very high distance accuracy of ~1 cm, and the camera offers images for object detection. However, the main problem of this system is its high cost. For example, a LiDAR with a large FoV and high distance accuracy, that is suitable for autonomous diving, costs about tens of thousands dollars. This high cost is one of the largest challenges for the system to be a dominant option for autonomous driving [e.g., 2].

Another option is adopting stereo cameras (e.g., Ambarella). Stereo cameras are able to detect objects using images and measure distances with a modest accuracy. Since a FoV of a single stereo camera is limited to about one hundred and twenty degrees (e.g., Stereolabs ZED 2), mainly due to the difficulty of accurate calibration, one possible solution is to place several cameras around a car in order to realize a large FoV. However, this makes the system expensive and complicated.

One possible solution for this problem is to broaden the FoV of a single stereo camera. In order for this, we have been developing an omnidirectional stereo camera with a novel optical system, whose design is presented in our previous studies [3, 4; see also 5–7]. This camera consists of two hyperbolic mirrors placed face-to-face, a wide-angle lens, and a single image sensor. This system satisfies all the requirements above. First, the hyperbolic mirrors enable a large FoV of 360-degree around. Second, since this is a stereo camera, images for object detection can be obtained simultaneously with distance information. Third, this camera exploits only a single set of a lens and sensor to obtain images from two viewpoints that are necessary for stereo vision, which saves the cost of a set of a lens and sensor. In addition, the hyperbolic mirrors are cost efficient compared to fisheye lenses as a way to realize the FoV of 360-degree around.

This optical system is designed to meet 5% distance error for objects 14 meters apart. This capability is adequate to locate objects, including other cars and pedestrians, in a crossroad when an autonomous car with this camera enters the crossroad. As another example, this detection range is large enough to perceive objects in three adjacent lanes for each side, which enables safe lane changes. Besides the short-range measurements described above, long-range measurements up to 300 meters are needed especially on highways. These are realized by using two of these cameras placed on two corners of a car separated by approximately 2 meters to construct a stereo camera with a longer baseline [3, 4]. This longer-baseline stereo camera is valid only for objects farther than ~5 m because of occlusion. Therefore, this optical system realizes seamless short-range omnidirectional measurements by a single camera and long-range measurements by combining two cameras, whose FoV and accuracy satisfy requirements of autonomous driving.

On the other hand, a new calibration method, where sensor images are corrected for distortion, also needs to be developed to obtain accurate distance information, because the optical system is totally novel and the FoV is much larger than ordinal perspective cameras. As an easy calibration method for ordinal cameras, Zhang [8] proposed a method using a checkerboard, which can be flexibly placed. Mei and Rives [9] expanded the Zhang's method to be applied to omnidirectional cameras including those using a hyperbolic mirror. Since our optical system is a combination of two hyperbolic mirrors and a wide-angle imaging lens, their method can basically be applied to our system. However, it does not deal with a wide-angle imaging lens and some of misalignments of the system. This insufficient calibration leads to inaccurate and even loss of distance information.

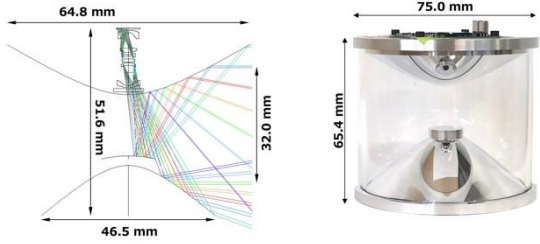

Figure 1. (*Left*) Optical design of the omnidirectional stereo camera. The color lines show the incident rays, which are reflected by the hyperbolic mirrors and enter the lens system. (*Right*) Constructed prototype following the optical design.

In this paper, we develop a new calibration method for our novel omnidirectional stereo camera, concentrating on the short-range configuration. Based on the Mei and Rives procedure, we newly take into consideration a fisheye-like wide-angle imaging lens, a tilt of the image sensor, and an offset between the mirrors and lens. Distance measurements are conducted using a single omnidirectional stereo camera. We evaluate the accuracies of the new calibration method and calculated distances.

The structure of this paper is as follows. In Section 2, we describe our omnidirectional stereo camera and a constructed prototype. In Section 3, we construct an optical model for the new calibration method. This calibration method is conducted to the prototype and distance information is calculated. In Section 4, the calibration accuracy and distance accuracy are evaluated, and we discuss our findings. We give a summary in Section 5.

## 2. The Optical Design and Constructed Prototype of the Omnidirectional Stereo Camera

Here we describe the optical design and prototype of our new omnidirectional stereo camera.

2.1. Optical Design

The optical design of the omnidirectional stereo camera is shown in the left panel of Figure 1. This system consists of two hyperbolic mirrors, a single image sensor, and a fisheye-like wide-angle lens unit. Rays incident toward focuses of the hyperbolic mirrors are reflected in the direction of the lens unit and imaged on the image sensor. This means that the rays reflected by the upper and lower mirrors compose upper- and lower-view images, respectively. The former rays, which reflected by the upper mirror, are imaged in the inner region of the sensor, and the latter, which reflected by the lower mirror, are imaged in the outer region. The lens unit is designed to reduce comatic aberration and astigmatism induced by the mirrors. This system is cost efficient because it uses hyperbolic mirrors rather than fisheye lenses to obtain 360-degree around images. Another reason is that it acquires upper- and lower-images by using only a single set of a lens and sensor.

As described in the previous section, the target distance error is 5% for objects 14 meters apart. This accuracy in turn corresponds to 5% distance error for objects 300 meters apart, when two of these cameras, separating by approximately 2 meters, are used.

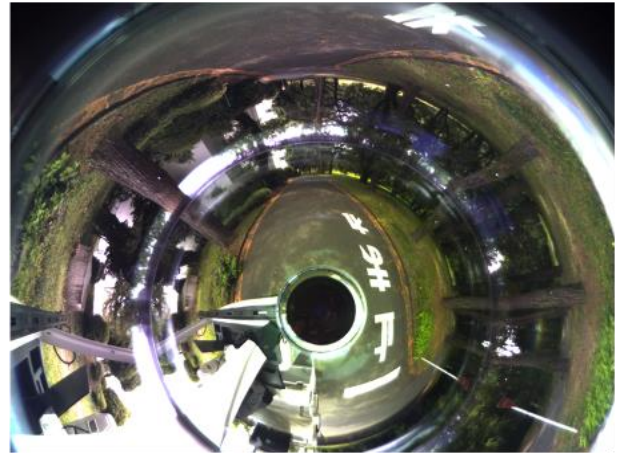

Figure 2. Sensor image taken by the prototype. The inner and outer regions correspond to the upper- and lower-view images, respectively.

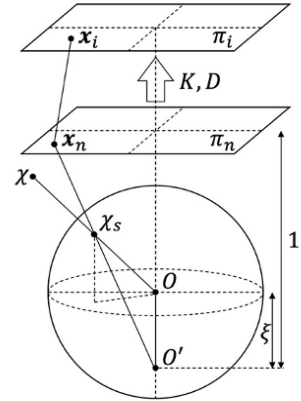

Figure 3. Optical model for the calibration. $K$ and $D$ represent a camera matrix and distortion coefficients, respectively.

2.2. Constructed Prototype

The constructed prototype is presented in the right panel of Figure 1. Besides the components described at the optical design, a glass cylinder is added to hold the upper and lower mirrors. We use an image sensor of IDS UI-3592LE-C Rev. 2, whose resolution is $4912 \times 3684$. Although this optical system can realize 360-degree FoV, a part of the FoV becomes ineffective when the camera is installed on one of the corners of a car. Thus, for this prototype, the image sensor is placed off axis to make the horizontal FoV 270 degrees. The vertical FoV of upper and lower images are from −50 degrees to +10 degrees and from −20 degrees to +10 degrees, respectively. Figure 2 shows a sensor image obtained by the prototype.

## 3. The New Calibration Method

In this section, we describe a new calibration method for our omnidirectional stereo camera. Based on the Mei and Rives procedure [9], we newly take into consideration a fisheye-like wide-angle imaging lens, a tilt of the image sensor, and an offset between the mirrors and lens.

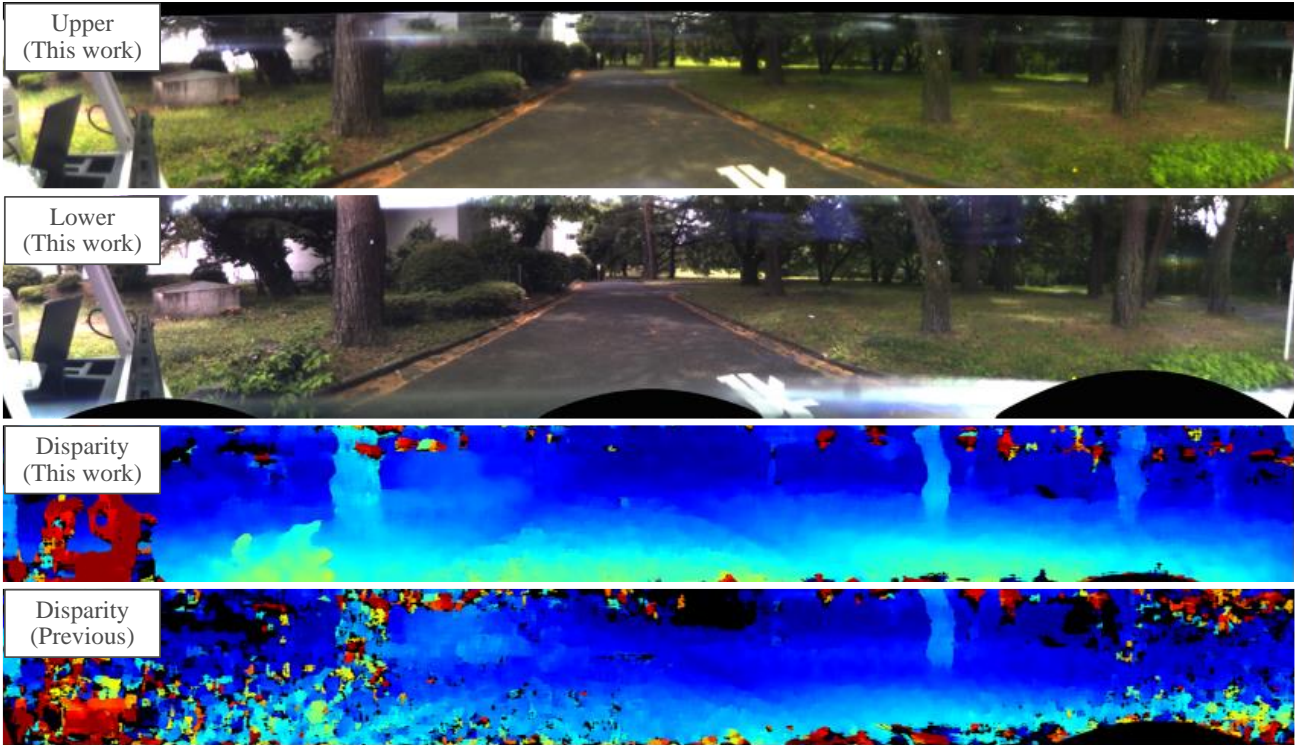

Figure 4. Calculated images using our and the previous models from the sensor image presented in Figure 2. (*Upper and upper middle*) Upper- and lower-view images cylindrically projected using our new optical model. (*Lower middle*) Disparity image calculated from the upper two images. Near and far objects are shown in red and blue, respectively. (*Lower*) Disparity image calculated from upper- and lower-view images calibrated using the previous model.

3.1. New Calibration Method

In the process of obtaining distance information from the sensor image, there are three steps. The first step is to expand the image. We adopt the cylindrical expansion [e.g., 10] because the image of the region we are focusing on are not significantly distorted with this projection. Another reason is that corresponding points in each image align in vertical lines for the short-range configuration, which makes it easy to apply traditional block-matching procedures. Simultaneously with the above cylindrical expansion, we remove image distortion. This is called a camera calibration procedure. The second step is image rectification, where the upper- and lower-view images are transformed to be projected onto the same cylinder. The last step is to conduct the block-matching procedure to calculate disparities between the two images.

For reliable distance measurements, an accurate calibration is required. Since our lens unit has a large distortion and the prototype can be affected by some misalignments, we need a new calibration method that considers these factors. In addition, the large FoV of this prototype makes an accurate calibration difficult.

Our method is based on a calibration method for omnidirectional cameras by Mei and Rives [9], which incorporates a flexible calibration method using a calibration board [8]. This method assumes a parametrical optical model to simulate the optical system. If the model is accurate enough, where distortion and misalignments are sufficiently reproduced, an accurate cylindrical expansion is also made possible.

In order to construct an accurate optical model, based on the Mei and Rives model (hereafter previous model), we consider additional five factors: higher-order radial distortion, radial dependence of tangential distortion, thin prism distortion, a tilt of the image sensor, and a lens-mirror offset. We describe the previous model and those five additional factors in detail below [see also 11, 12, 13].

*The previous model.* Figure 3 illustrates the previous model. It consists of a unit sphere, normalized plane $\pi_n$, and image plane $\pi_i$. The object point $\chi$ is projected onto the unit sphere ($\chi_s$) and reprojected onto $\pi_n$ after changing the origin $O$ to $O'$ by the amount of $\xi$. This process emulates the reflection by the hyperbolic mirror. Then the point $\boldsymbol{x_n}$ on $\pi_n$ is converted to $\boldsymbol{x_i}$ on pixel coordinates considering lens effects including radial distortion and tangential distortion.

*Additional higher-order radial distortion.* Since our lens unit has a fisheye-like wide view angle, higher-order radial distortion needs to be considered. Sixteen-order distortion is implemented while the previous model uses only four-order distortion,

$$\begin{bmatrix} \delta x_i \\ \delta y_i \end{bmatrix} = \begin{bmatrix} x_n \\ y_n \end{bmatrix}(k_1 r^2 + k_2 r^4 + k_3 r^6 + k_4 r^8 + k_5 r^{10} + k_6 r^{12} + k_7 r^{14} + k_8 r^{16}), \quad (1)$$

where $r = \sqrt{x_n^2 + y_n^2}$.

*Additional detailed tangential distortion.* In order to correct for misalignments of the lens, more detailed components for tangential distortion need to be considered. Radial dependence of tangential distortion ($q_1 - q_3$) is added to the previous components ($p_1, p_2$),

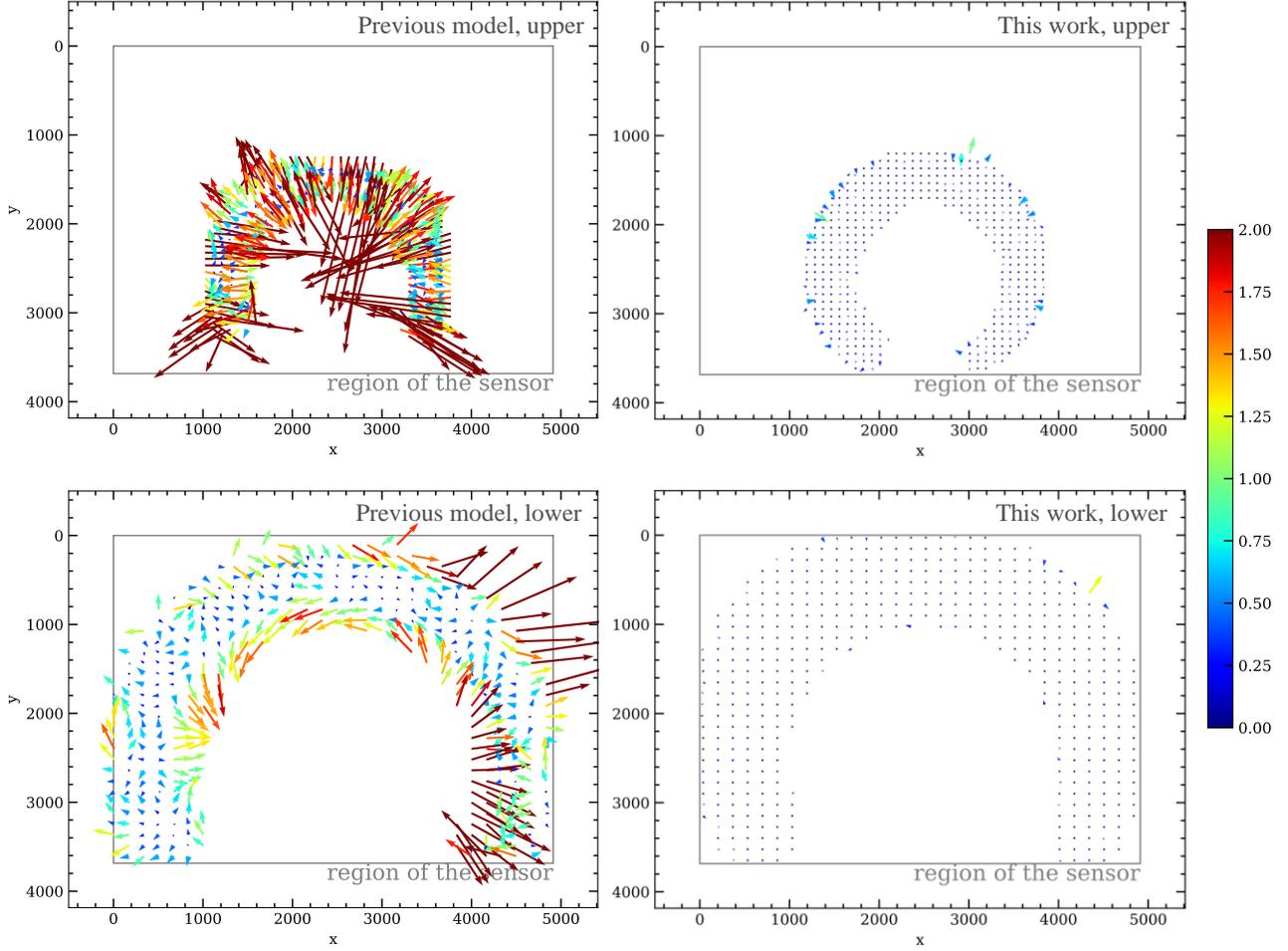

Figure 5. Binned differences between the image-detected and model-predicted circle positions for upper- (*upper*) and lower- (*lower*) view images. The left and right panels show the results of the previous model and our new optical model, respectively. The vectors show the position differences in pixel unit but emphasized 200 times and the colormap shows the amount of the differences in pixel unit. The region of the sensor is shown with a gray rectangular.

$$\begin{aligned}\delta x_i &= \left(2p_1 x_n y_n + p_2(r^2 + 2x_n^2)\right)\\ &\quad (1 + q_1 r^2 + q_2 r^4 + q_3 r^6)\\ \delta y_i &= \left(p_1(r^2 + 2y_n^2) + 2p_2 x_n y_n\right)\\ &\quad (1 + q_1 r^2 + q_2 r^4 + q_3 r^6).\end{aligned} \quad (2)$$

*Additional thin prism distortion.* A small amount of tilt of lenses causes thin prism distortion. This distortion can be described by the following components ($s_1 - s_4$),

$$\begin{aligned}\delta x_i &= s_1 r^2 + s_2 r^4\\ \delta y_i &= s_3 r^2 + s_4 r^4.\end{aligned} \quad (3)$$

*Additional sensor tilt.* Since the image sensor can be tilted, we take this effect into consideration by tilting the image plane. Positions of the projected point on the tilted image plane is

$$s \begin{bmatrix} x_{i,t} \\ y_{i,t} \\ 1 \end{bmatrix} = \begin{bmatrix} R_{33}(\tau_x, \tau_y) & 0 & -R_{13}(\tau_x, \tau_y) \\ 0 & R_{33}(\tau_x, \tau_y) & -R_{23}(\tau_x, \tau_y) \\ 0 & 0 & 1 \end{bmatrix} R(\tau_x, \tau_y) \begin{bmatrix} x_i \\ y_i \\ 1 \end{bmatrix}, \quad (4)$$

where $s$ is an arbitrary scale factor, $\tau_x, \tau_y$ are angle parameters, and

$$R(\tau_x, \tau_y) = \begin{bmatrix} \cos(\tau_y) & 0 & -\sin(\tau_y) \\ 0 & 1 & 0 \\ \sin(\tau_y) & 0 & \cos(\tau_y) \end{bmatrix} \begin{bmatrix} 1 & 0 & 0 \\ 0 & -\cos(\tau_x) & \sin(\tau_x) \\ 0 & \sin(\tau_x) & \cos(\tau_x) \end{bmatrix}. \quad (5)$$

*Additional lens-mirror offset.* In the previous model, it is assumed that the axes of the hyperbolic mirror and lens are aligned. However, there can be an offset between them. In order to model the offset, we add offset parameters to the project point $x_n$, which in turn changes the origin of lens distortion.

$$\begin{aligned}\delta x_n &= \Delta x_n \\ \delta y_n &= \Delta y_n\end{aligned} \quad (6)$$

As a result, our new model has 27 parameters, while the previous model has ten parameters including camera parameters. On the other hand, the increase of the number of the parameters requires more computation time and more calibration images to reliably constrain the parameters. This time, we used hundreds of calibration images.

Simultaneously with the calibration procedure described above, we can obtain the external parameters of the calibration boards. We use the external parameters of the boards detected in both the views for the image rectification procedure so that the two coordinate systems align.

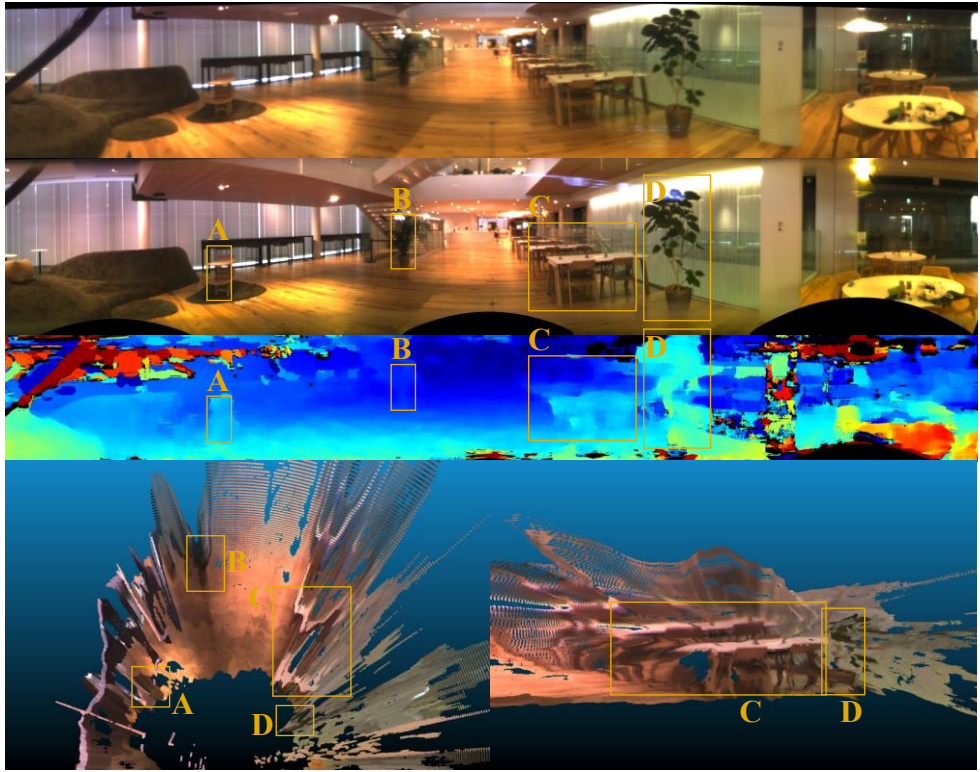

Figure 6. RGB images, disparity image, and point cloud of inside scene. For clarity, we show only 180 degrees of the FoV. A, B, C, and D shows the corresponding objects. Among the four tables of the object C, the nearest and farthest are placed 4.5 meters and 12 meters apart from the camera, respectively.

### 3.2. Result of the New Calibration Method

We conduct an optimization of the parameters in the same manner as the omnidir::calibrate function implemented in the OpenCV (Open Source Computer Vision) library. As the calibration board, we use a circle-grid board, which performs better than a checkerboard [14]. The optimized parameters are shown in Tables 1 and 2. The results of the cylindrical projections are shown in the upper two panels of Figure 4. The disparity image is calculated using cv::stereoBM function implemented in the OpenCV library and shown in the lower middle panel of Figure 4. For a comparison, we also present a disparity image calculated using the previous model in the lower panel of Figure 4. More disparity and clear gradient of the ground are obtained by our calibration method.

## 4. Discussion

In this section, we first discuss the accuracy of our calibration method. Then, we evaluate the distance accuracy.

### 4.1. Evaluations of the Calibration Error

If an optical model accurately reproduces the camera, the calibration by the optical model is also accurate. One of the methods to evaluate the calibration accuracy is to evaluate the accuracy of the model on the sensor image. We calculate the differences between the image-detected and model-predicted circle positions of calibration boards (Figure 5). The length and the direction of a vector represent the amount and the direction of the systematic error of the model at each point. Although there remains systematic errors when the previous model is used, our model significantly reduces the errors.

The root mean squares of the differences by the previous model are 1.67 pixels and 1.12 pixels for upper- and lower-view images, respectively, which have non-negligible effects on the distance accuracy. On the other hand, using our optical model, the root mean squares improve from 1.67 pixels to 0.28 pixel and from 1.12 pixels to 0.26 pixel for upper- and lower-view images, respectively. These are 6.0 and 4.3 times improvements for upper- and lower-view images, respectively.

### 4.2. Obtained Disparity Images and Point Clouds

In order to present the accuracy and limitation of our camera, we show the RGB images, disparity images, and point clouds in Figures 6 and 7. In Figure 6, images and point clouds taken inside are presented. Objects up to 14 meters apart including the flat floor, plants, wooden shelf, and tables are reproduced. Figure 7 shows images and point clouds taken outside. Within a range of 14 meters, structures of the paved road, trees, and curbs are reproduced. This result proves that this camera satisfies the requirement for autonomous cars to detect objects, such as pedestrians, other cars, and obstacles within the range.

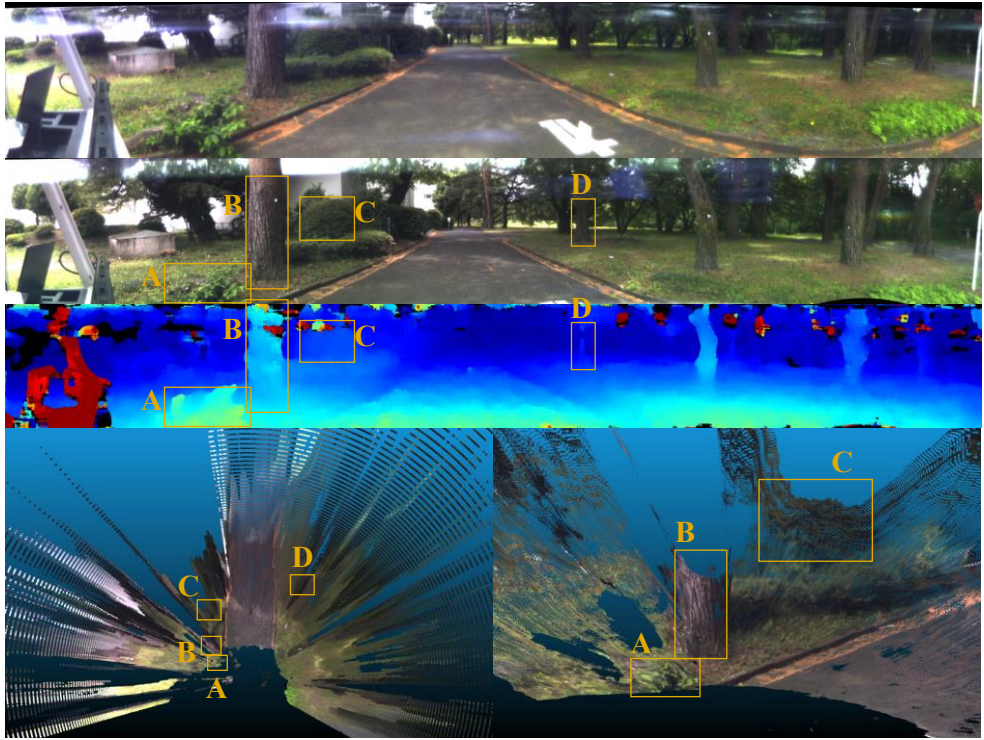

Figure 7. Same as Figure 6 but for outside scene.

### 4.3. Evaluations of the Distance Error

We evaluate the distance accuracy using an obtained point cloud shown in Figure 6. In this point cloud, a flat wooden floor is reproduced. By comparing measured and actual distances of every single point belonging to the floor, we evaluate measurement deviations from the true value. Figure 8 shows the measured distances against actual distances to the points. The 1 $\sigma$ standard deviations are at most 4.9% for points up to 14 meters apart. The systematic error is 5.7% for all the range. Compared to those values using the previous model, the 1 $\sigma$ standard deviations improves from ~10% to 4.9%, the systematic error improves from ~43% to 5.7%, which is an almost nine times improvement. Considering several errors including those by image acquisition, this prototype performs as planned, because the systematic error of 5.7% is virtually identical to the target value of 5%. On the other hand, the optical resolution of the prototype is degraded compared to its best performance by eight times in spot size. This can also degrade the standard deviation and systematic error. This standard deviation can be reduced inversely proportional to the square root of the number of pixels, when several pixels that belongs to the same object are taken into consideration.

### 5. Conclusion

We have developed a new calibration method for our omnidirectional stereo camera in order to accurately correct for the distortion to accomplish 5% distance error for objects 14 m apart by the short-range configuration. Based on the previous calibration method by Mei and Rives, we have improved the calibration accuracy by considering higher-order radial distortion, detailed tangential distortion, an image sensor tilt, and a lens-mirror offset.

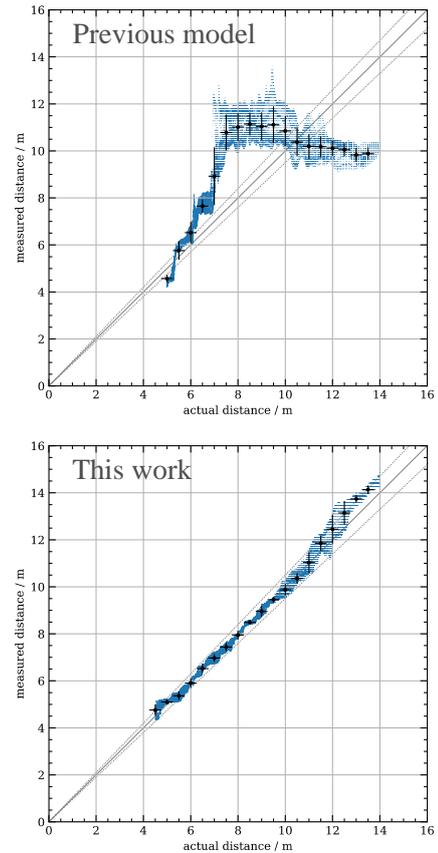

Figure 8. Measured distance against actual distance between 4.5 and 14 m. The blue points show each point from a part of point cloud shown in Figure 6. The black points represent the average in the bin and the vertical error bars 1 $\sigma$ standard deviation. The bold gray line represents an accurate distance measurement and thin gray lines show 5% deviation from the true value.

Our method reduces the root mean square of the calibration accuracy by 6.0 times and 4.3 times for upper- and lower-view images, respectively. The systematic error of distance is 5.7% up to objects 14 meters apart, which is improved almost nine times compared to that calculated by the previous method and virtually identical to the target value of 5%. The standard deviation is at most 4.9%. As future work, an improvement in the optical resolution can lead to further improvement in accuracy.

We have demonstrated that our camera is capable of locating objects, such as pedestrians, other cars, and obstacles within 14 meters. Besides distance measurements described in this paper, object recognition is also needed for autonomous driving. Further improvements and optimizations to our tentative recognition system will make this camera practical as a cost-effective sensor for autonomous driving.

This paper is written based on a proceeding presented at JSAE 2021 Annual Congress.

Table 1. Best-fit parameters of the optical model for the upper view

| | New optical model | OpenCV optical model | Description |
|---|---|---|---|
| **Camera parameters** | | | |
| $f_x, f_y$ | 907.2, 908.2 | 1295.1, 1295.2 | Focal lengths in x- and y-directions in pixel units |
| $c_x, c_y$ | 2486.4, 2669.1 | 2443.5, 2601.4 | Principal points in x- and y-directions |
| $s$ | −4.3768 | −1.1024 | Skew coefficients |
| **Distortion coefficients** | | | |
| $k_1, k_2, k_3, k_4, k_5, k_6, k_7, k_8$ | −2.0914×10⁻², 1.4286×10⁻¹, −7.2762×10⁻², 1.4879×10⁻², 1.3799×10⁻³, −1.1389×10⁻³, 1.8171×10⁻⁴, −9.7028×10⁻⁶ | −1.6360×10⁻¹, −4.5147×10⁻¹, 0, 0, 0, 0, 0, 0 | Radial distortion coefficients |
| $p_1, p_2$ | 1.0159×10⁻², 1.1788×10⁻² | −3.7040×10⁻³, −5.5740×10⁻³ | Tangential distortion coefficients |
| $q_1, q_2, q_3$ | −1.7712×10⁻¹, 3.5490×10⁻², −3.1020×10⁻³ | 0, 0, 0 | Radial dependencies of the tangential distortion |
| $s_1, s_2, s_3, s_4$ | −1.2406×10⁻², 6.3520×10⁻⁴, −1.4512×10⁻², 8.7730×10⁻⁴ | 0, 0, 0, 0 | Thin prism distortion coefficients |
| **Other parameters** | | | |
| $\xi, \xi', \Delta x, \Delta y$ | 0.57583, 0, −6.8619×10⁻², −8.5941×10⁻² | 1.2256, 0, 0, 0 | Mirror parameters and mirror−lens offset coefficients |
| $\tau_x, \tau_y$ | −5.7219×10⁻², 6.8473×10⁻² | 0, 0 | Tilted sensor coefficients |

Table 2. Best-fit parameters of the optical model for the lower view

| | New optical model | OpenCV optical model | Description |
|---|---|---|---|
| **Camera parameters** | | | |
| $f_x, f_y$ | 1716.3, 1747.0 | 1801.8, 1805.2 | Focal lengths in x- and y-directions in pixel units |
| $c_x, c_y$ | 2721.7, 2635.1 | 2437.6, 2594.3 | Principal points in x- and y-directions |
| $s$ | 0 | −1.6857 | Skew coefficients |
| **Distortion coefficients** | | | |
| $k_1, k_2, k_3, k_4, k_5, k_6$ | −1.4137×10⁻¹, 1.3832×10⁻¹, −1.7294×10⁻¹, 1.1855×10⁻¹, −4.6525×10⁻², 1.0562×10⁻², −1.2914×10⁻³, 6.5830×10⁻⁵ | −1.1063×10⁻¹, 8.7920×10⁻³, 0, 0, 0, 0, 0, 0 | Radial distortion coefficients |
| $p_1, p_2$ | 1.5136×10⁻², 1.4125×10⁻² | −6.8752×10⁻⁴, −4.8280×10⁻⁴ | Tangential distortion coefficients |
| $q_1, q_2, q_3$ | 1.1063×10⁻¹, −6.5861×10⁻², 7.2110×10⁻³ | 0, 0, 0 | Radial dependencies of the tangential distortion |
| $s_1, s_2, s_3, s_4$ | −1.9775×10⁻², 1.4820×10⁻³, −8.4950×10⁻³, 6.7358×10⁻⁴ | 0, 0, 0, 0 | Thin prism distortion coefficients |
| **Other parameters** | | | |
| $\xi, \xi', \Delta x, \Delta y$ | 0.77735, 0, −1.8999×10⁻¹, −1.7602×10⁻² | 0.78967, 0, 0, 0 | Mirror parameters and mirror−lens offset coefficients |
| $\tau_x, \tau_y$ | 7.2186×10⁻³, 1.3330×10⁻¹ | 0, 0 | Tilted sensor coefficients |